\documentclass[manuscript]{acmart}

\setcopyright{none} 
 \renewcommand\footnotetextcopyrightpermission[1]{}
\AtBeginDocument{%
  }

\usepackage{multirow}
\usepackage{subcaption}
\usepackage{makecell}
\usepackage{xcolor}
\usepackage{comment}


\begin{document}


\title{Who (or what) are they prompting? How CS students conceptualize and use LLMs}


 \author{Nikitha Donekal Chandrashekar}
\affiliation{%
  \institution{Virginia Tech}
  \city{Blacksburg}
  \country{USA}}
\email{nikitha@vt.edu}

\author{Margaret Ellis}
\affiliation{%
  \institution{Virginia Tech}
  \city{Blacksburg}
  \country{USA}}
\email{maellis1@vt.edu}

\author{Sehrish Basir Nizamani}
\affiliation{%
  \institution{Virginia Tech}
  \city{Blacksburg}
  \country{USA}}
\email{sehrishbasir@vt.edu}

\author{Naren Ramakrishnan}
\affiliation{%
  \institution{Virginia Tech}
  \city{Alexandria}
  \country{USA}}
\email{naren@vt.edu}

\begin{abstract}

Large language models (LLMs) are increasingly used by students in computing courses.
Yet computing education research has not fully examined how teaching and inclusion of LLMs shape students' conceptualization and understanding of LLMs, and how these factors relate to their subsequent use. 
Although prior work has examined prompting strategies and AI literacy, the relationships among students' conceptualizations of LLMs, their understanding of LLM mechanisms, and their prompting behavior remain underexplored. 
We studied these relationships in a post CS1 course that introduces students to the various CS subfields. 
We adapted the course to teach and allow the usage of LLMs. 
We administered an end-of-course survey, using open-ended items to elicit students' representations of LLMs and their explanations of how LLMs work, alongside closed-response measures of prompting style and prompt count ($N=283$ students consented and provided codable responses).
Using qualitative coding, we categorized students' understanding of how LLMs work into four ordinal knowledge levels based on the Structure of Observed Learning Outcomes (SOLO) taxonomy and coded their metaphoric conceptualizations along two dimensions: persona type and relational role. 
Our results show that students most often conceptualized LLMs as human assistants, and context-rich prompting was the preferred style.
The study contributes a two-dimensional characterization of students’ LLM metaphoric representation and exploratory analyses also identified associations among these constructs showing how understanding these systems relates to their usage.


\end{abstract}


  
\maketitle

\section{Introduction and Motivation}
\label{sec:intro}

Educators have been scrambling for strategies to adjust to the rampant availability and use of Large Language Models (LLMs), the technology which underpins modern Artificial Intelligence (AI). 
Questions arise about how to prepare students for an AI-infused workforce and still emphasize fundamentals and critical thinking skills. 
The need for this balance is accentuated in computer science education due to the tech industry’s rapid and well-publicized adoption of AI \cite{dora2025genai,raihan2025sigcsellmlitreview}.

Computer Science (CS) students are likely to eventually have layered levels of interaction with LLMs as users, integrators, and future creators of AI.
Yet, students are often using AI chatbots or coding agents without understanding how they work.  
Familiarity with the foundations of LLMs such as transformer model, tokens, attention mechanism, training, and inference has the potential to not only prepare students for future careers, but affect how they use/prompt AI themselves.

\begin{figure*}[!htpb]
  \centering

  \includegraphics[width= 1.0\linewidth]{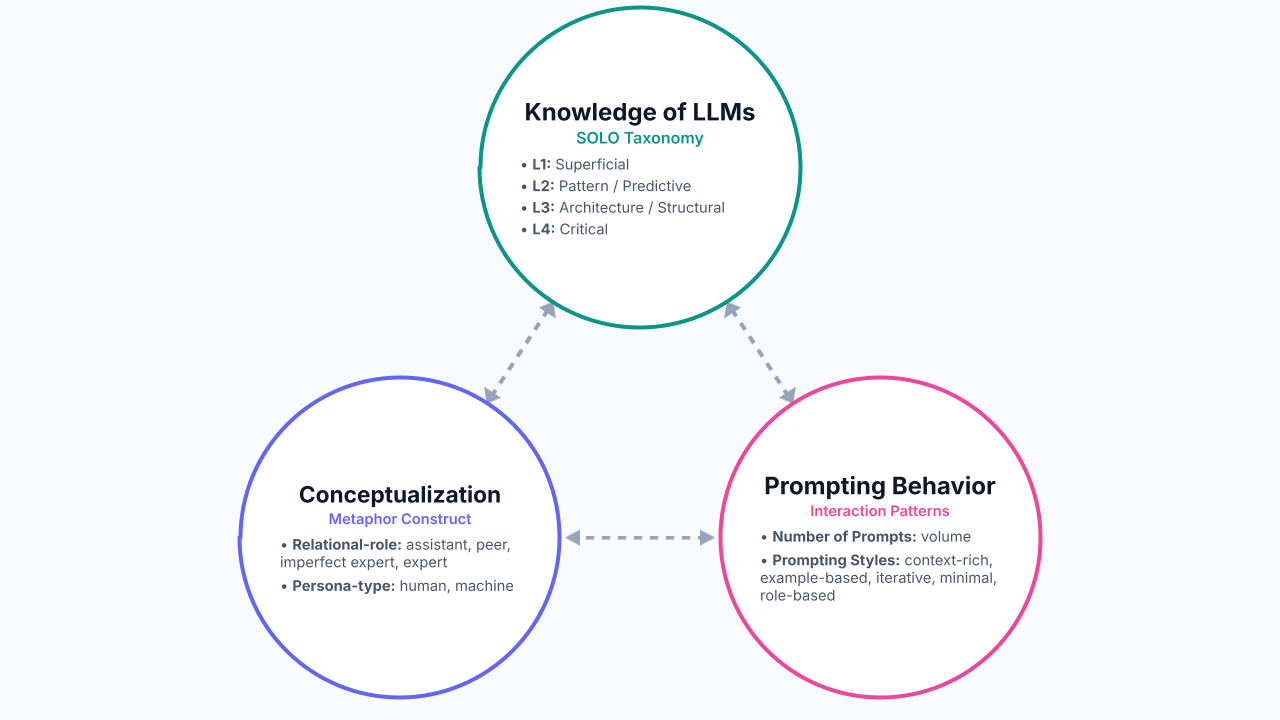}
  \caption{Study constructs and relationships.}

  \Description{Diagram connecting LLM Knowledge to LLM Conceptualization and Prompting Behavior, and also LLM Conceptualization to Prompting Behavior. The Knowledge of LLMs SOLO Taxonomy is broken down L1 Superficial, L2 Pattern/Predictive, L3 Architectural/Structural, L4 Critical and Prompting Behavior. The Conceptualization metaphor constructs are broken down into relational-role(assistant, peer, imperfect expert, expert) and persona-type(human, machine) facets. The Prompting Behavior interaction patterns are number of prompts and prompting styles(context-rich,example-based,iterative,minimal,and role-based). }
  \label{fig:hypoth}
\end{figure*}

We integrated lessons to demystify how LLMs work into our CS Problem Solving Course where computer science majors are introduced to several subfields of computer science such as software engineering, human computer interaction, networking, and data science \cite{donekal2026demystify}.
In this research paper, we study the relationships between the students' conceptualization of LLM, LLM knowledge-level, and reported prompting as shown in \autoref{fig:hypoth}. 
We organized these constructs into three hypotheses to guide our research:

\begin{itemize}
  \item \textbf{H1.} Students' conceptualization of LLMs is associated with their knowledge of how LLMs work.
  \item \textbf{H2.} Students' conceptualization of LLMs is associated with their prompting behavior, operationalized as prompting style and prompt count.
  \item \textbf{H3.} Students' knowledge of how LLMs work is associated with their prompting behavior.
\end{itemize}

We collect data through surveys with two open-ended and two closed questions that were administered at the end of the semester.
We studied how students completed the metaphor ``An LLM is like a \_\_\_\_\_\_\_'' to understand their conceptualization of LLMs. 
We also analyzed how students explained how LLMs worked according to the Structure of the Observed Learning Outcomes (SOLO) taxonomy ~\cite{biggs1982evaluating}. 

Our main contributions are the two-dimensional metaphor coding scheme and the descriptive characterization of students’ LLM knowledge, conceptualization, and reported prompting.
Findings of this study are that LLM knowledge level was significantly associated with students'reported use of context-rich prompting, and students who framed LLMs as a human persona type used role-based prompting and had a higher prompt count.

\section{Related Work}
\label{sec:related-work}

In 2023 Lau and Guo~\cite{lau2023ban} closed their empirical study of instructor attitudes toward AI coding tools with open research questions, among them: "What mental models do novices currently form both about the code that AI generates and about how the AI works to produce that code?"~\cite{lau2023ban} 
Our study takes a step toward answering this and how these relate to prompting behavior in the Problem Solving in CS course. 
Here we discuss related work on prompting, LLM conceptualization(metaphors), and LLM knowledge.

Some CS education research has investigated  student prompting behavior. 
Amoozadeh et al.~\cite{amoozadeh2024student} found that students adopted a variety of prompting and problem-solving strategies when working with ChatGPT and that frequency of AI use alone did not predict success.
Penney et al.~\cite{penney2025outcomes} found that students provided context-rich prompts to online human tutors, but not LLMs, in a study of the interactions of 20 novice programmers. 
When working with a human tutor, students spontaneously provided background, framed questions strategically, and iterated far more richly.
One of the conclusions of this study was that students needed to be taught AI literacy and prompt engineering.

After studying student interaction and performance with a CodeTutor chatbot in 2024, Lyu et al. determined that quality student prompts that were clear and detailed improved effectiveness with CodeTutor. 
They concluded that students need more support for Generative AI literacy \cite{Lyu2024EvaluatingTE}. 
We propose that understanding how LLMs work, and not just prompting strategies, improves student effectiveness with AI. 
To our knowledge, whether students' understanding of how LLMs work predicts their prompting behavior has not been examined.

In the Science article \textit{Metaphors of Artificial Intelligence}, author Melanie Mitchell references the various metaphors used to conceptualize AI, such as role players, cultural technologies, and autocomplete on steroids ~\cite{mitchellscience2024metaphors}. 
She highlights how the design of recent systems causes users to anthropomorphize AI since it uses first person pronouns and expresses feelings, and she argues that the impression of AI having a "mind" affects how users trust and build relationships with it. 
Our work investigates the relation between student LLM metaphors and their understanding of how LLMs work as well as their prompting behavior.

Two computer science education studies explored how students conceptualize LLMs, but did not provide specific ties to their prompting behavior.
Zhou et al.~\cite{zhou2024ethics} interviewed 20 computing education stakeholders (undergraduates, master students, PhD students, faculty, and industry practitioners)  and identified three mental models held by computing education stakeholders: Writing Tool, Coding Tool, and Information Tool, each associated with distinct use patterns and ethical concerns but all of which included the general precaution to use careful, systematic prompts. 
Smith et al.~\cite{smith2024early} document a wider range of student framings across learning, coding, and writing use cases, from ``calculator for coding'' to parallel colleague, suggesting that students possess a variety of LLM conceptualizations. However, the authors do not identify generalized relational roles, such as tool or peer, for AI or connect these framings to prompting.

Budhiraja et al.~\cite{budhiraja2024jarvis} collected data from multiple Indian Engineering Universities with surveys from 480 CS students and interviews from 17 CS students. 
They organized student interactions around three relational roles: Supervisor for Programming, Brainstorming Companion, and Tutor. 
They also explored student impressions of their prompting experience which included the need to set context. 
However, they do not connect prompting to the perceived relational roles. 
Although prior work has separately studied what students do when they prompt and, how they conceptualize LLMs, little is known about whether and how students’ conceptualizations of LLMs are associated with their prompting behavior.

The need for AI Literacy is widely recognized, with the U.S. Department of Labor releasing an AI Literacy Framework in 2026 ~\cite{DOL2026AILiteracy}.
Among several outcomes, this framework mentions understanding both AI's core concepts and how to provide the right context to AI.
And the Digital Education Council provides a five dimension framework for AI literacy in higher education, each at three competency levels: 1) Understanding AI and Data, 2) Critical Thinking and Judgment, 3) Ethical and Responsible Use, 4) Human-Centricity, Emotional Intelligence, and Creativity, and 5) Domain Expertise ~\cite{DEC_AILiteracyFramework_2025} which emphasizes critical engagement and is geared more towards verification and ethics than prompting strategies. 

Although these three areas have been independently investigated, our study focuses on the relationships between how students understand, conceptualize, and prompt AI.  

\section{Course Description}

Our institution is a large public university situated in the South-east of the United States. 
In the Fall of 2025 this post-CS1 Problem Solving course integrated current computer science tools such as version control and the command-line, and provided an introduction to LLMs and incorporated their usages throughout introductory units on algorithms, software engineering, web development, HCI, networking and data science. 
The course was taught across five sections with a total of 360 students and is required for CS majors.
The unit on LLMs introduced transformers and how system prompts and external data can improve responses. 
The limitations of LLMs were also highlighted along with emphasis on responsible use such as verification of outputs and awareness of privacy, safety, and fairness issues.

\section{Methods}
\label{sec:methods}

This study examines how students conceptualize LLMs, how those conceptualizations relate to their knowledge of how LLMs work, and how both relate to their prompting behavior.
Prompting behavior was operationalized through two self-reported measures: prompting-style and range of weekly prompt-count.
We administered a survey at the end of the course, coded the open-ended responses using established taxonomies, and tested for associations among the resulting variables. 
The remaining subsections describe the data collection procedure, the coding, and statistical analyses.
The statistical analyses are framed around the three hypotheses as described in \autoref{sec:intro}.
This study was exempted by the university IRB.

\subsection{Data Collection}

\paragraph{Survey instrument.}
The survey comprised two open-ended and two closed-response questions \autoref{tab:survey}.
The open ended questions focused on gathering data for conceptualization and knowledge of LLMs.
The closed responses focused on the prompting behavior of the students (count and style). 
The options for prompting style were derived from coding of open-ended responses in a previous semester \cite{donekal2026demystify}.

\begin{table}[t]
\centering
\caption{Survey Instrument Questions}
\label{tab:survey}
\small
\begin{tabular}{@{}p{2cm}p{2.7cm}p{2.8cm}@{}}
\toprule
Survey Item & Question & Response\\
\midrule
Conceptualization&
An LLM is like a \_\_\_\_\_\_\_ . & Open-Ended Response\\

Knowledge & How do LLMs work? & Open-Ended Response
\\
Prompting  & Approximately how many prompts do you submit to an LLM in a typical week? & Closed Response with five options: 0, 1--15, 16--50, 51--100, 100+\\

Prompting & When using LLM for learning, which prompting strategy do you most frequently use?&Closed Response for five prompting styles (minimal, context-rich, iterative, example-based, role-based) on a five point scale.\\
\addlinespace
\bottomrule
\end{tabular}
\end{table}

\paragraph{Participants.}
A total of $N=288$ (out of 360 enrolled students) students consented and completed the survey. 
Five students did not provide a codable response to the knowledge item, yielding $N=283$ responses for final analyses.

\subsection{Analysis}

The analysis of the data was conducted in two phases. The first phase focused on qualitative coding of the open-ended response of the students.
In the next phase of the analysis, we conducted statistical tests to validate the hypotheses guiding the research.

\subsubsection{Qualitative Coding.}

The knowledge item was coded deductively using the SOLO taxonomy ~\cite{biggs1982evaluating}, which characterizes understanding as progressing from disconnected, surface observations to integrated and abstract reasoning. 
We mapped this progression onto a four-level rubric of students' understanding of LLMs.

The metaphor scheme used a combined deductive–inductive process: initial codes were drawn from prior work and refined against the present responses.
We began with three codes drawn from prior work~\cite{donekal2026demystify}: Assistant, Peer, and Expert. Because a substantial number of responses did not fit cleanly within these three codes, the authors iteratively refined the scheme. 
This process produced two distinct coding dimensions. 
The first, \emph{ontological persona representation}, captures whether a student frames the LLM as a human or machine. 
The second, \emph{relational role}, captures the relationship or status the student ascribes to the LLM, expanding the original three codes to four: Assistant, Peer, Imperfect Expert, and Expert. 
Treating persona and role as separate dimensions allowed a single metaphor response to be characterized along both axes.
The final code description used for coding are detailed in the next section.

Because code definitions were iteratively refined during coding to ensure they were applied consistently across all responses, inter-rater reliability was not calculated. 
Disagreements were resolved through consensus, and all co-authors reviewed and accepted the final codes.


\begin{table}[t]
\caption{Coding schemes for LLM conceptualization.}
\label{tab:codesconcept}
\centering
\begin{tabular}{p{2cm}p{1.8cm}p{3.6cm}}
\toprule
Dimension & Code & Description \\
\midrule

Persona Type &
Human &
Framed as a person, human role, or social actor. \\

&
Machine &
Framed as a tool, information system, or technological artifact. \\


Relational Role &
Assistant &
Supportive technology or helper. \\

&
Peer &
Collaborative, equal-status partner. \\

&
Imp. Expert &
Knowledgeable but fallible guide. \\

&
Expert &
Authoritative source of knowledge. \\

\bottomrule
\end{tabular}
\end{table}

\subsubsection{Code Descriptions}
The descriptions of the codes for the two dimensions of the LLM conceptualization are mention in \autoref{tab:codesconcept}.
The codes for both relational-role and persona dimensions are categorical.
\autoref{tab:codesknowledge} presents the descriptions of the codes for LLM knowledge level and their corresponding SOLO Level. 
The codes for knowledge levels are ordered from L1 (Superficial) to L4 (Critical).
These code descriptions guide the coding.
   
\begin{table}[t]
\caption{Coding schemes for knowledge level.}
\label{tab:codesknowledge}
\centering
\begin{tabular}{p{2.2cm}p{1.8cm}p{3.4cm}}
\toprule
Code & SOLO Level & Conceptualization \\
\midrule
L1: Superficial &
Pre-/Uni-structural &
Treats the LLM as a thinking human or search engine. \\

L2: Pattern/Pred. &
Multi-structural &
Understands the LLM as a probabilistic next-word predictor. \\

L3: Arch./Struct. &
Relational &
Mentions transformers, tokens, or training processes. \\

L4: Critical &
Extended Abstract &
Recognizes limitations such as hallucination, alignment, or optimization. \\
\bottomrule
\end{tabular}
\end{table}

\subsubsection{Statistical analysis.}
We conducted two types of statistical analyses to examine the associations specified in our hypotheses. 
For H1 and H2, the predictor variables were categorical and the outcome variables were ordinal. 
For H1, we examined whether students’ conceptualizations of LLMs, analyzed through persona type and relational role, were associated with ordered knowledge levels (L1 $<$ L2 $<$ L3 $<$ L4). 
For H2, we examined whether the same conceptualization variables were associated with ordinal prompting outcomes, including prompting-style ratings and prompt count. 
We used the Mann-Whitney U test for the binary persona-type dimension and the Kruskal-Wallis test for the four-category relational-role dimension. 
Significant or marginally significant Kruskal-Wallis results were followed by Dunn’s post-hoc tests with Holm correction for multiple comparisons.

For H3, both variables were ordinal: knowledge was coded as an ordered level, and each prompting outcome was measured on an ordered response scale. 
We therefore used Spearman’s rank correlation to examine associations between knowledge and each prompting outcome. 
To compare students’ ratings across the five within-participant prompting styles, we used the Friedman test.
We treated p < .05 as statistically significant.
Analyses were conducted in Python using the SciPy and scikit-posthocs libraries.
The authors also report the use of LLM (Claude code) to help write the scripts for analysis.


\section{Results}
\label{sec:results}

This section reports the findings in four stages. 
We first describe the distributions of the student responses for the three constructs: LLM conceptualization, LLM knowledge and  prompting behavior.
We then evaluate the three hypotheses: whether LLM conceptualization is associated with knowledge (H1), whether LLM conceptualization is associated with prompting behavior (H2), and whether knowledge is associated with prompting behavior (H3). 

\subsection{Descriptive Statistics}

Participants ($N=283$) were substantially more likely to frame AI in human terms than in machine terms.
As shown in \autoref{tab:desc}, 192 (68\%) participants used a human-persona framing (e.g "smart friend", "tutor"), compared with 91 (32\%) who used a machine-persona framing (e.g. "a tool", "instant Google").
Knowledge of LLMs was concentrated in the middle of the ordered knowledge scale.
Pattern/Predictive knowledge was the modal category ($n=143$), followed by Architecture/Structural ($n=76$) and Superficial ($n=60$).
The Critical category contained only four participants and is therefore not sufficiently populated to support independent inferential claims. The most common relational role was Assistant (e.g."helper", "unpaid intern") and the least common was Imperfect Expert (e.g. "biased mentor", "knower of all and master of none").

\begin{table}[t]
\caption{Distribution of coded categories and representative responses.}
\label{tab:desc}
\centering
\begin{tabular}{llr}
\toprule
Dimension & Codes & $n$ \\
\midrule

\multirow[t]{2}{*}{Persona Type}
    & Human & 192 \\
    & Machine & 91 \\


\multirow[t]{4}{*}{Relational Role}
    & Assistant & 152 \\
    & Peer & 56 \\
    & Imperfect Expert & 12 \\
    & Expert & 63 \\


\multirow[t]{4}{*}{LLM Knowledge}
    & Superficial & 60 \\
    & Pattern/Predictive & 143 \\
    & Architecture/Structural & 76 \\
    & Critical & 4 \\

\bottomrule
\end{tabular}
\end{table}

Prompt counts were low and tightly distributed: 82\% of participants reported between 1 and 50 prompts (the 1--15 and 16--50 bands), with a median band of 16--50.
The five prompting styles differed significantly across participants' ratings (Friedman $\chi^2(4)=193.4$, $p<.001$).
On the 0--4 use scale, where 2 indicates ``Moderately my style,'' context-rich prompting received the highest mean ($M=2.49$), followed by example-based ($M=2.36$), iterative ($M=2.18$), and minimal ($M=1.69$), with role-based prompting received the lowest mean ($M=1.24$), a difference exceeding one scale point from context-rich prompting. 
This ordering was also evident when each participant's highest-rated style was treated as their dominant style: context-rich prompting was dominant for 118 participants, minimal prompting for 99 participants, and the remaining three styles for substantially fewer participants.

\subsection{H1: Association between LLM Conceptualization and LLM Knowledge}

H1 predicted that participants' conceptualization of LLMs would be associated with their knowledge of how LLMs work.
We treated knowledge as an ordered outcome, Superficial scored as 1 to Critical scored as 4, and tested each conceptualization dimension against it: persona type, which has two levels, and relational role, which has four. 
For persona type, we conduct Mann-Whitney $U$ test and for relational role, we conduct Kruskal--Wallis test.
We additionally perform a Dunn's post-hoc (Holm corrected) analysis to identify which specific roles differed. 
Table~\ref{tab:h1} summarizes the results.

Persona type showed a small association with knowledge ($p=.036$, $r=.14$; $p_{\mathrm{adj}}=.065$). 
Participants who framed LLMs as machines reported higher knowledge ($M=2.22$) than those who framed them as human ($M=2.02$).
Relational role was not significant at the .05 level ($\varepsilon^2=.015$, $p=.065$). 
In Dunn's post-hoc (Holm-corrected), Assistant roles reported higher knowledge than Peer roles (2.16 vs.\ 1.86, $p=.044$).

Taken together, these results provide limited empirical support for H1. 
The interpretation of these associations is complicated by dependence between the two dimensions of LLM conceptualization. 
Students characterizing LLMs as Peers reported the lowest knowledge and also predominantly used human metaphors (53 of 56). 
The lower knowledge of the human-persona group is therefore largely the same participants as the low-knowledge Peer group, so the persona-type difference cannot be cleanly separated from the role difference. 
We therefore interpret H1 as receiving only limited, exploratory support: the association is small, and cannot be localized cleanly to either dimension.



\begin{table}[t]
\caption{Mean LLM knowledge score by conceptualization category and their corresponding p-values from running the Mann-Whitney (MW), Kruskal-Wallis (KW) tests. *: p-value $<$ 0.05.}
\label{tab:h1}
\centering
\begin{tabular}{p{2 cm}p{2.2 cm}p{1cm}p{1.2 cm}}
\toprule
Dimension & Code & Mean Score & KW/MW (p-value) \\
\midrule
Persona Type
    & Human & 2.02& \textbf{0.036*} \\
    & Machine & 2.22 \\
Relational Role
    & Assistant & {2.16} & 0.065 \\
    & Peer & {1.86} \\
    & Imperfect Expert & 2.08 \\
    & Expert & 2.10 \\
\bottomrule
\end{tabular}
\end{table}

\subsection{H2: Associations between LLM Conceptualization and Prompting Behavior}

H2 states that conceptualization of LLMs would be associated with prompting
behavior. 
We tested each of the two conceptualization dimensions against six prompting
outcomes: the five prompting styles and prompt count.
Similar to previous hypothesis, for persona type, we conduct Mann-Whitney $U$ test and for relational role, we conduct Kruskal--Wallis test.
Two small associations were found to be statistically significant, and both involved persona type.

First, persona type was associated with role-based prompting. 
Participants who framed AI as human used role-based prompting more than those who framed it as a machine (1.34 vs.\ 1.04; $p=.027$, $r=.16$; $p_{\mathrm{adj}}=.23$).
Second, persona type was associated with prompt count: human-persona participants reported more prompts than machine-persona participants (median band 16--50 vs.\ 1--15; $p=.038$, $r=.14$; $p_{\mathrm{adj}}=.23$).

Beyond these two, the Kruskal--Wallis test also showed that relational role was only marginally associated with context-rich prompting ($\varepsilon^2=.015$, $p=.067$). 
Students who conceptualized LLMs as an Expert reported higher use of context-rich prompting ($M=2.73$) and it was lowest among those who framed LLMs as an Imperfect Expert ($M=2.00$), with Assistant ($M=2.51$) and Peer ($M=2.27$) in between. 
However, Dunn's post-hoc test (Holm-corrected) revealed no significant pairwise differences (all $p>.13$).

The remaining prompting outcomes, namely minimal, iterative, and example-based prompting, were not significantly associated with either conceptualization dimension. 
Hence, H2 received partial, exploratory support: two small associations reached nominal significance, both involving persona type (human framing accompanied more role-based prompting and higher prompt volume), but neither survived correction for multiple comparisons ($p_{\mathrm{adj}}=.23$).




\subsection{H3: Association between LLM Knowledge and Prompting Behavior}

H3 states that knowledge about how LLMs work may be associated with the prompting behavior. 
Since, both variables here are ordinal: knowledge runs from Superficial to Critical, and each prompting style is rated on an ordered scale from ``Not at all my style'' to ``Very strongly my style'', the appropriate test for two ordered variables is therefore a rank correlation, which asks whether the two tend to increase together.
We used Spearman's $\rho$ to relate knowledge to each of the six prompting outcomes.

Only context-rich prompting was associated with knowledge, and the correlation was weak ($\rho=.15$, 95\% CI $[.03,.26]$, $p=.013$; $p_{\mathrm{adj}}=.078$), accounting for about 2\% of the variance ($\rho^2=.02$). 
No other style, nor prompt count, showed a significant association. 
The correlation is positive, meaning knowledge and context-rich prompting preference rise together: higher-knowledge participants lean more toward context-rich prompting on average. 
Average context-rich reporting rises across the three well-populated knowledge levels, from $2.25$ at Superficial to $2.48$ at Pattern/Predictive to $2.72$ at Architecture/Structural.

H3 therefore received partial, exploratory support. 
Higher LLM knowledge showed a small association with greater preference for context-rich prompting. 
LLM knowledge was unrelated to the other styles or to prompt volume.



\subsection{Summary}
Overall, students conceptualized LLMs as human rather than machine, and most often as an assistant.
Most students described LLMs at the level of next-word prediction (Pattern/Predictive), with fewer articulating the underlying architecture.
Several exploratory associations were observed from our analysis.
First, LLM conceptualization was related to their knowledge of how LLMs work (H1), where students framing LLMs as machines had higher LLM knowledge than students framing LLMs as humans.
Second, conceptualization of LLMs was more reliably tied to prompting behavior (H2), where participants who framed the LLM as human used role-based prompting and reported higher prompt volume.
Thirdly, knowledge about LLMs was related to a single prompting style (H3), where higher knowledge accompanied greater use of context-rich prompting.
The association of the persona type of LLM conceptualization to prompting behavior is an interesting and important outcome from our study that would be investigated in detail in our future works. 

\section{Discussion and Threats to Validity}

This study is motivated by an interest in understanding students’ conceptualizations of AI and how computing courses can support students' effective use of these systems.
Our results suggest that students with better understanding of how LLMs work were more likely to frame them as machines rather than as human-like entities and more likely to report using context-rich prompting. 
These patterns suggest that an understanding of how AI works may be associated with ways of reasoning about and interacting with AI, particularly when students recognize that AI outputs depend on the context, constraints, and information provided in the prompt. 
At the same time, students who conceptualized LLMs using a human persona type reported greater use of role-based prompting and higher prompt counts.
This pattern suggests that anthropomorphic framings may encourage more conversational or role-oriented interactions with AI, even when such framings do not necessarily reflect stronger understanding of how these systems work.
These findings motivate further investigation into whether instruction about LLM mechanisms can promote effective AI use.



Students in this study reported more complex prompting strategies than those observed in some earlier CS1 studies, where students often defaulted to minimal or low-effort queries~\cite{penney2025outcomes, amoozadeh2024student}. 
The relatively high use of context-rich prompting in this study is therefore a finding worth examining further. 
This pattern may reflect the nature of the course content, which explicitly introduced LLMs and incorporated their use across multiple computing topics. 
It may also reflect broader changes in students’ familiarity with generative AI tools over time, or it may indicate that this population of post-CS1 students had more prior experience with LLMs beyond their coursework. 
The post-course survey alone is not sufficient to attribute these differences to the specific effects of the course content on students’ LLM knowledge, conceptualization, or prompting behavior.

One threat to validity of this study is that the data is from self-reported surveys which are vulnerable to response bias, e.g. students may not have accurate impressions or reporting of their prompting style. Also, this study is only one set of students in a specific course, at a specific institution, so the results may not be generalizable, but are a first step in that direction. 
Qualitative coding can also introduce interpretive bias that may affect outcomes. 

\section{Future Work}
Future research includes investigating changes in student LLM conceptualization and knowledge throughout a course and the direct analysis of student interactions with LLMs via coding agents.
The researchers aim to expand the study to incorporate related implications of LLM knowledge and conceptualization, such as verification and ethical decision-making practices, which could also inform instructional approaches. 
Furthermore, we are interested in understanding the relationships between these constructs across a broader population of students.

\section{Conclusion}
This study characterizes post-CS1 students’ LLM conceptualizations, knowledge, and reported prompting. 
Students most often framed LLMs as human assistants, and context-rich prompting was the highest-rated style.
Exploratory associations among these constructs emerged, where students' conceptualization of LLMs is associated with their knowledge of how LLMs work and also with their prompting behavior when considering persona dimension.
LLM knowledge level was also associated with students’ use of context-rich prompting and framing of AI as an assistant.
These patterns motivate future pre-post work examining whether instruction about LLM mechanisms is associated with changes in students’ use of AI.

\bibliographystyle{ACM-Reference-Format}
\bibliography{prompting_LLM}


\end{document}